# Heterostructures of MXene and N-doped graphene as highly active bifunctional electrocatalysts

Si Zhou, Xiaowei Yang, Wei Pei, Nanshu Liu, Jijun Zhao[*]

*Key Laboratory of Materials Modification by Laser, Ion and Electron Beams (Dalian University of Technology), Ministry of Education, Dalian 116024, China*

**Abstract**

MXenes with versatile chemistry and superior electrical conductivity are prevalent candidate materials for energy storage and catalysts. Inspired by recent experiments of hybridizing MXenes with carbon materials, here we theoretically design a series of heterostructures of N-doped graphene supported by MXene monolayers as bifunctional electrocatalysts for oxygen reduction reaction (ORR) and hydrogen evolution reaction (HER). Our first-principles calculations show that the graphitic sheet on $V_2C$ and $Mo_2C$ MXenes are highly active with ORR overpotential down to 0.36 V and reaction free energies for HER approaching zero, both with low kinetic barriers. Such outstanding catalytic activities originate from the electronic coupling between the graphitic sheet and MXene, and can be correlated to the $p_z$ band center of surface carbon atoms and the work function of the heterostructures. Our findings screen a novel form of highly active electrocatalysts by taking advantage of the fast charge transfer kinetics and strong interfacial coupling of MXenes, and illuminate a

---

universal mechanism for modulating the catalytic properties of two-dimensional hybrid materials.

**Introduction**

Two-dimensional (2D) early transition metal carbides, nitrides and carbonitrides, known as MXenes, constitute a large family of 2D materials and recently attract enormous attentions.[1] These monolayers are generally produced by extracting the A element from the MAX phases (M is early transition metal; A is group IIIA or IVA element; X is C or N).[2] To date, 19 different MXenes, such as $Ti_3C_2$, $Mo_2C$, $V_2C$, $Ti_4N_3$ and $Ta_4C_3$, have been synthesized in laboratory.[3-6] These MXenes are good electrical conductors with high elastic moduli.[7-10] The rich chemistries and unique morphologies render MXenes versatile for sensors,[11] energy storage,[12, 13] catalysis,[14, 15] and water purification.[16] In particular, the layered structures of MXenes allow the storage and rapid transport of ions, and hence are widely exploited for electrodes of supercapacitors[13] and metal-ion batteries.[17, 18] Some theoretical studies predicted that the MXene surfaces terminated by oxygen functional groups are active for catalysis of hydrogen evolution reaction (HER)[19-21] and $CO_2$ reduction.[22, 23] However, so far only $Mo_2C$ monolayer has been demonstrated to have HER activity in experiment.[15]

On the other aspect, MXenes are excellent conducting reinforcement to composites, showing strong interfacial coupling and fast charge transfer kinetics.[9, 24, 25] Thus, integration of MXenes can effectively enhance the electrochemical properties of composite materials.[26-28] Geng *et al*. synthesized large-area $Mo_2C$ MXene on graphene template.[29] The heterostructure is active for HER electrocatalysis with onset voltage much lower than the $Mo_2C$-only electrodes. Wu *et al*. fabricated 2D hierarchical nanohybrids by assembling few layer $MoS_2$ nanoplates on the $Ti_3C_2$

MXene backbone, and obtained HER activity competitive to that of the $MoS_2$-based catalysts.[25] Zhao *et al.* hybridized 2D metal-organic frameworks (MOF) with $Ti_3C_2$ MXene nanosheets and observed prominent activity for oxygen evolution reaction (OER).[30]

Electrochemical reactions including HER, OER, as well as oxygen reduction reaction (ORR) are the cornerstones of many renewable energy devices such as fuel cells, metal air batteries and water electrolysis.[31-33] The large-scale application of these technologies relies on development of active, stable and low-cost catalysts to replace the scarce noble metal catalysts. Benefited from the synergic effects, the composites of N-doped graphitic carbon and transition metal based materials form one main category of electrocatalysts for HER, OER and ORR.[31, 33, 34] Considering the metallic nature of MXenes and their efficient charge transfer kinetics, it is intriguing whether the hybrids of MXenes and N-doped carbon materials can serve as a new family of catalysts with even superior performance.

Here we explore the electrocatalytic properties of the heterostructures of N-doped graphene supported by MXenes — $Ti_2C$, $V_2C$, $Nb_2C$ and $Mo_2C$ monolayers. First-principles calculations demonstrate that $V_2C$ and $Mo_2C$ hybridized with N-doped graphene are highly active for both HER and ORR with small overpotentials and low kinetic barriers. Some pivotal issues behind the remarkable catalytic properties are addressed: How are the electronic band structure and surface properties of graphene impacted by the MXene substrate? What are the factors mediating the coupling strength between graphene and MXene? What are the key parameters fundamentally determining the catalytic activity of these heterostructures? Our theoretical explorations elucidate the synergic effect of the graphene/MXene hybrids from atomistic level, and reveal the key parameters toward rational design of multifunctional carbon/MXene based hybrid electrocatalysts.

**Methods**

Density functional theory (DFT) calculations were performed by the Vienna ab initio simulation package (VASP),[35] using the planewave basis set with an energy cutoff of 550 eV, the projector augmented wave (PAW) potentials,[36] and the GGA-PBE exchange-correlation functional.[37] To model the graphene/MXene heterostructures, we used a supercell consisting of $5 \times 5$ graphene unit cells and $4 \times 4$ unit cells for $Ti_2C$, $Nb_2C$, $Mo_2C$, and $\sqrt{19} \times \sqrt{19}$ unit cells for $V_2C$, respectively, giving lattice mismatch below 2.12% (see Table S1† for details). We considered O functional groups terminating the bottom surface of MXene, as the $Ti_2C$, $V_2C$, $Nb_2C$ and $Mo_2C$ MXenes terminated by O species are thermodynamically more stable than those by OH- or F-termination.[21] A vacuum region of 16 Å was applied in the vertical direction. For these hybrid systems, the in-plane lattices of MXenes were either stretched or compressed to fit that of graphene. The strain effect on the binding property and catalytic activity is found to be negligible (Table S2†). The graphene sheet was then substitutionally doped by graphitic (2.0 at. %) or pyridinic (6.1 at. %) N atoms, which are the typical doping concentrations of synthetic N-doped graphitic carbon catalysts.[38, 39] For each hybrid system, we constructed four models with the graphitic layer placed on different positions relative to the underneath MXene sheet (the four models exhibit similar structural, electronic and catalytic properties, as demonstrated by Table S1, S3-6†). The Brillouin zone was sampled by $3 \times 3 \times 1$ uniform **k** point mesh. Within the constrained supercells, the model structures were fully optimized using thresholds for the total energy of $10^{-4}$ eV and force of 0.02 eV/Å, respectively. The Grimme's DFT-D3 scheme of dispersion correction was adopted to describe the van der Waals (vdW) interactions in these layered systems.[40] Partial charge densities were evaluated by the Bader charge analysis.[41] Within the current slab model, work function was computed by referring the Fermi energy to the electrostatic potential in vacuum.

The ORR overpotentials ($\eta^{ORR}$) were calculated by the standard hydrogen electrode (SHE) method,[42] considering the four-electron reaction pathway in alkaline media:[43]

$$* + O_2 (g) + H_2O (l) + e^- \rightarrow OOH^* + OH^- \quad (1)$$

$$OOH^* + e^- \rightarrow O^* + OH^- \quad (2)$$

$$O^* + H_2O (l) + e^- \rightarrow OH^* + OH^- \quad (3)$$

$$OH^* + e^- \rightarrow * + OH^- \quad (4)$$

where * represents an adsorption site on the catalyst surface; OOH*, O* and OH* are the oxygen intermediates. The Gibbs free energy of formation ($\Delta G_i$, $i = 1, 2, 3, 4$) was computed for each ORR step (see Equation S1-6† for details). The overpotential $\eta^{ORR}$ is then given by:

$$\eta^{ORR} = \Delta G^{ORR}/e + U_0 \quad (8)$$

where $\Delta G^{ORR}$ = max[$\Delta G_1$, $\Delta G_2$, $\Delta G_3$, $\Delta G_4$], and $U_0$ = 0.40 V is the equilibrium potential for pH = 14 and temperature $T$ = 298 K.[44] The computed $\eta^{ORR}$ is actually independent on the pH value.[45]

The HER performance was characterized by the reaction free energy ($\Delta G_{H*}$) of hydrogen adsorption ($\Delta G_{H*}$), defined as:[46]

$$\Delta G_{H*} = \Delta E_{H*} + \Delta ZPE - T\Delta S \qquad (5)$$

where $\Delta E_{H*}$, $\Delta ZPE$ and $\Delta S$ are the differences of DFT total energy, zero-point energy, and entropy between the adsorb H* phase and $H_2$ gas phase, respectively; $\Delta ZPE$ and $\Delta S$ were acquired by vibrational frequency calculation (Table S7†). Following this methodology, $\eta^{ORR}$ and $\Delta G_{H*}$ can be obtained by computing the binding energies of relevant reaction intermediates on various sites of the catalyst surface. Here we define binding energy as the energy of the adsorbed reaction intermediate relative to the energies of $H_2$ and $H_2O$ molecules (Equation S9-12†). Note that $\eta^{ORR}$ and $\Delta G_{H*}$ are prerequisite parameters characterizing the electrocatalytic properties of materials for ORR and HER, respectively. Based on these parameters and the methodology described above, the trends of ORR and HER activities for a variety of catalytic materials, such as transition metals (compounds),[46, 47] carbon materials[48, 49] and 2D transition metal dichalcogenides,[15, 50] have been successfully predicted.

The kinetic barriers and transition states for the ORR and HER reactions were simulated by the climbing-image nudged elastic band (CI-NEB) method.[51] Five images were used to mimic the reaction path. The intermediate images were relaxed until the perpendicular forces were smaller than 0.02 eV/Å.

**Results and Discussion**

N-doped graphene is known to be capable of catalyzing ORR and HER with the active sites originated from defects or edges.[52] To activate the carbon basal plane, we hybridize N-doped graphene with the metallic $Ti_2C$, $V_2C$, $Nb_2C$ and $Mo_2C$ monolayers, as shown in Fig. 1 (hereafter denoted as $G/Ti_2C$, $G/V_2C$, $G/Nb_2C$ and $G/Mo_2C$, respectively). The detailed structural information of these graphene/MXene models is given by Table 1, Table S1† and Fig. S1-4.† Synthesis of such heterostructures would be feasible, as recent experiments directly grow $Mo_2C$ MXene films on graphene[29] and $Mo_2C$ nanoparticles on graphene nanoribbons[53] by chemical vapor deposition.

The graphene/MXene heterostructures show interlayer distances of 2.13~2.40 Å and binding energies of −0.42~−0.17 per C atom in graphene. As a result of strong interfacial coupling, prominent electron transfer occurs from MXene to the graphitic sheet, with each C atom in graphene gaining 0.06~0.11 $e$ from the underneath metal atoms. These transferred electrons would fill the C $p_z$ orbitals and disturb the π conjugation of graphene. Consequently, reactivity of the carbon surface of heterostructures is greatly improved. The oxygen intermediates are strongly adsorbed on the graphene/MXene hybrids with binding energies much lower than those on freestanding N-doped graphene (Table S3-6†). The overall oxygen binding strength follows the sequence: $G/V_2C$ > $G/Mo_2C$ > $G/Nb_2C$ > $G/Ti_2C$, with OH* binding energies in the range of −0.19~0.84 eV, 0.40~0.95 eV, 0.50~1.13 eV and 0.58~1.32 eV, respectively (Fig. S5†). Moreover, the binding energies of OH*, OOH* and O* species show linear relations with each other, as those observed for the transition metal based materials[47, 54] and graphitic carbon materials.[44, 55] Such linear correlation leads to a volcano relationship between the catalytic activity and binding energies of key reaction intermediates, which connects high activity to moderate binding strength — known as the Sabatier principle.[47, 56]

Fig. 2a, b plot the activity volcano of ORR using the OH* binding energy as a descriptor. Most of the C sites on G/V$_2$C are located on the left side of volcano. The oxygen binding is relatively strong, such that desorption of OH* species to form an OH$^-$ anion is difficult and limits the reaction rate of ORR. For G/Mo$_2$C, G/Nb$_2$C and G/Ti$_2$C, on the other hand, most C sites are on the right side of volcano. The oxygen binding strength is relatively weak, and hence dissociation of an O$_2$ molecule to form OOH* species is the rate-limit step. The origins of the overpotentials for various graphene/MXene heterostructures are clearly illustrated by the free energy diagrams in Fig. 2c, d.

The binding energies and ORR overpotentials of the graphene/MXene hybrids highly depend on the structural environment of surface C atoms, i.e., the position relative to N dopants and the underneath MXene. Due to the electron transfer from MXene to surface C atoms and from C to N atoms, the graphitic sheet of the heterostructures is associated with a non-uniform electron density distribution. Generally, the C atoms with less electron densities provide larger binding strength with oxygen intermediates (Fig. S6†).[44] The strongest binding is achieved on the C atoms close to both N atom and the hollow site of topmost metal atoms in MXene. In contrast, the C atoms close to the top site of MXene form chemical bonds with the underneath metal atoms and gain more electrons than those on the hollow site, and thus they bind weakly with oxygen intermediates. In particular, G/V$_2$C and G/Mo$_2$C contribute a large number of active sites at the summit of the volcano, all of which coming from the hollow-site C atoms. The most active sites are the hollow-site C atoms close to the pyridinic N dopants, yielding the lowest overpotentials of 0.36 and 0.39 V for G/V$_2$C and G/Mo$_2$C, respectively. The G/Nb$_2$C and G/Ti$_2$C systems provide weaker binding strength and have ORR overpotentials above 0.54 and 0.64 V, respectively. Note that the overpotentials of these heterostructures are much lower than that of freestanding N-doped graphene (1.24 V), and rather competitive to the benchmark Pt catalyst (0.45 V[57]) as well as the hybrid catalysts of N-doped graphene and Co or Fe metals (0.41 and 0.38 V, respectively, calculated by using the same method[44]). The detailed catalytic properties of the graphene/MXene models are given by Table S3-6† and Fig. S1-4.†

To further evaluate the catalytic performance, we examine the reaction barriers for an O$_2$ molecule to dissociate on the active sites of G/V$_2$C and G/Mo$_2$C, which may be the key step limiting the kinetics of ORR. According to our NEB calculations, an O$_2$ molecule can efficiently dissociate on G/V$_2$C and G/Mo$_2$C through dual reaction pathways: it either reacts with a H$_2$O molecule to form an OOH* group, or directly dissociate into two O* species, as illustrated by Fig. 3. Benefited from the strong binding capabilities of these heterostructures, both reaction pathways are exothermic and kinetically readily occur. The barriers of the $O_2 \rightarrow OOH^*$ and $O_2 \rightarrow 2O^*$ pathways are 0.23 and 0.20 eV for G/V$_2$C, and 0.68 and 0.76 eV for G/Mo$_2$C, respectively (Fig. S7-8†). In particular, the $O_2 \rightarrow 2O^*$ pathway is thermodynamically favorable, which can promotes the active sites on the right side of volcano in Fig. 2b. For those C sites, formation of OOH* from $O_2$ is the rate-limit step and gives rise to the largest potential step, as displayed in Fig. 2c, d. This potential step can be avoided by direct dissociation of $O_2$ into O* species, such that the entire ORR process is only limited by the kinetic barriers.

The enhanced surface reactivity of the graphene/MXene heterostructures is also beneficial for HER catalysis. The C atoms having strong binding with oxygen intermediates also favor the adsorption of H* species, as governed by the linear relations between the binding energies of various reaction intermediates (Fig. S5†). As shown in Fig. 4a, G/V$_2$C and G/Mo$_2$C provide moderate binding strength for H* adsorption, with adsorption free energy nearly in equilibrium with that of gaseous H$_2$. Specifically, the C atoms close to the hollow site of V$_2$C substrate have $\Delta G_{H^*}$ of −0.04~0.17 eV and are eligible for HER catalysis. The most active sites with $\Delta G_{H^*} =$ −0.04 and 0.04 eV come from the hollow-site C atoms close to the pyridinic and graphitic N dopants, respectively, quite competitive to Pt ($\Delta G_{H^*} = -0.10$ eV from our calculations). For G/Mo$_2$C that is less reactive than G/V$_2$C, only a few hollow-site C atoms close to the N dopants can provide sufficient binding strength for HER catalysis with $\Delta G_{H^*}$ of 0.05~0.26 eV. For G/Nb$_2$C and G/Ti$_2$C, the H* binding is too weak for HER catalysis with $\Delta G_{H^*} > 0.2$ eV.

The kinetic process of $H_2$ evolution through a Tafel mechanism[58] is then investigated for the graphene/MXene heterostructures, in which two H* species desorb and form a $H_2$ molecule. The reaction barriers are about 1.33 and 1.56 eV for G/V$_2$C and G/Mo$_2$C, respectively (Fig. 4b, Fig. S9†), moderately higher than that of Pt (0.8 eV[59]) and close to the values of MoS$_2$ edges (1.0~1.5 eV[58, 60]). Note that the reaction barrier is usually much lower under the Heyrovsky mechanism that involves the combination of H* species with a proton accompanied by electron transfer.[20, 58] Therefore, HER may proceed even more facilely through the Volmer-Heyrovsky pathway[58] for the graphene/MXene heterostructures.

Note that standalone Ti$_2$C, V$_2$C, Nb$_2$C and Mo$_2$C MXene monolayers terminated by O species were predicted to have HER activity at hydrogen coverage of 12.5%~50%.[15, 19, 20] When hybridized with N-doped graphene, the bottom surface of Ti$_2$C, V$_2$C and Mo$_2$C substrates retains HER activity, while the bottom surface of Nb$_2$C substrate binds too weakly with H* species and is inactive for HER. Therefore, both the graphitic sheet and the bottom surface of MXene substrate in G/V$_2$C and G/Mo$_2$C can provide active sites for HER catalysis. Due to the weak binding with oxygen intermediates, the O-terminated MXene surfaces (either standalone or hybridized ones) do not show ORR activity with overpotentials above 1.6 V (see Table S8† for details).

As the synthetic MXenes may have F residues on the surface,[1] we examine their impact on the electrocatalytic properties of the graphene/MXene heterostructures. Due to the strong electron-withdraw ability of F atoms, the binding capability of the heterostructures with MXene (taking $V_2C$ as representative) fully terminated by F atoms is reduced compared to the O-terminated ones, with binding energies of reaction intermediates raised by about 0.2 eV (Table S9†). As a result, the data points in the ORR activity volcano shift to higher binding energies. The $G/V_2C$ and $G/Mo_2C$ hybrids can still provide reaction sites for ORR catalysis with activity comparable to that of the O-terminated systems. As the H* binding strength is also weakened, the presence of F species on MXene surfaces would be adverse to the HER activity of the heterostructures.

We also considered O-terminated $M_3C_2$ (M = Ti, V, Nb and Mo) MXenes to hybridize with N-doped graphene. Their surface binding properties and catalytic activities are quite similar to those of the $G/M_2C$ systems (Table S10†). Thus, the composites of 2D metal carbides and N-doped graphene may form one category of efficient electrocatalysts. Moreover, 2D metal nitrides have different surface polarity from that of metal carbides and may lead to diverse catalytic properties when hybridized with N-doped graphene, which would be the subject of our future study.

The synergic effect of the graphene/MXene hybrids can be understood from their electronic structures. All these hybrid systems show metallic behaviors, evident from the density of states (DOS) (Fig. S10-12†). The graphitic sheet is strongly hybridized with the MXene substrate, presenting prominent DOS at the Fermi level (Fig. 5a). Noticeably, the binding energies of reaction intermediates on the graphene/MXene heterostructures is correlated to the $p_z$ band center ($\varepsilon_{pz}$) of the graphitic sheet, as demonstrated by Table 1 and Table S1†. The binding strength decreases as $\varepsilon_{pz}$ approaches the Fermi level. Intuitively, the distinct C $p_z$ band centers reflect the electronic coupling strength between the graphitic sheet and MXene, that is, how the carbon bands are variated and shifted with respect to the Fermi level by the substrate interaction. Previous studies showed that the band shift of graphene on transition metal substrate depends on the work function of metals and interfacial potential step induced by the graphene-substrate interaction.[61] For our graphene/MXene systems with strong interfacial interaction (interlayer distance < ~2.4 Å), the potential steps are similar for the four types of heterostructures. MXenes with larger work function shift the graphene bands to deeper energies with respect to the Fermi level, as revealed by Fig. S11-12. A linear relation between $\varepsilon_{pz}$ and work function can be established: the graphitic sheet supported by MXene with a larger work function has lower $\varepsilon_{pz}$ and stronger binding capability (Fig. 5b and Fig. S13†). Therefore, for the graphene/MXene heterostructures, work function can be used as a descriptor, which can be directly modulated in experiment to tune the C $p_z$ band center and ultimately optimize the surface binding properties and catalytic activities.

The correlation between the binding strength and C $p_z$ band center can be interpreted by the local DOS of reaction intermediates adsorbed on the heterostructures (Fig. 5c). Two distinct peaks at about −10 and −5 eV correspond to the bonding and antibonding states formed between the C $p_z$ orbitals of the graphitic sheet and the adsorbate valence orbitals, respectively.[62, 63] In contrast to the transition metals with open $d$ shell, the antibonding states from the graphene/MXene hybrids are almost fully occupied, as there are very little electron states available in the conduction bands. As a result, deeper valence orbital levels of the graphitic sheet lead to larger binding strength with reaction intermediates according to the extended Hückel theory (see ESI† for details).[64]

Eventually, a profound insight into the design principles of the graphene/MXene hybrid electrocatalysts is gained: the surface binding property and catalytic activity are dictated by the electronic coupling between the graphitic sheet and MXene as well as the surface charge redistribution induced by N doping. By choosing appropriate MXene, it is feasible to modulate the surface binding capability toward the catalysis of specific electrochemical reactions. Furthermore, doping heteroatoms into the graphitic sheet can induce variations in the reactivity of C atoms near and far from the dopants, offering an effective strategy for designing bifunctional catalysts. As a new form of 2D hybrid catalysts, the graphene/MXene heterostructures may possess not only high activities, superior conductivities and tunable functionalities but also large surface area and excellent mechanical properties, therefore promising for flexible and portable energy devices.

**Conclusion**

In summary, we theoretically design 2D heterostructures of N-doped graphene supported by MXene monolayers as highly active electrocatalysts. Our first-principles calculations show that the graphitic sheets hybridized with $V_2C$ and $Mo_2C$ MXene monolayers exhibit remarkable catalytic activities for both ORR and HER. The ORR overpotential is as low as 0.36 V involving kinetic barrier of only 0.2 eV. The HER process has $\Delta G_{H*}$ approaching zero and Tafel reaction barrier down to 1.3 eV. The favorable activities of these heterostructures are attributed to the strong electronic coupling between the graphitic sheet and MXene, which alters not only the graphene band profile but also the band center relative to the Fermi level. The catalytic activity can be correlated to the C $p_z$ band center of the graphitic sheet and the work function of the heterostructures. These vital understandings help prescribe the principles of compositing MXenes and carbon materials as a novel form of multifunctional hybrid electrocatalysts. These exciting results would trigger experimental and theoretical efforts into the innovative carbon/MXene composites for flexible and portable energy devices.

**ACKNOWLEDGMENTS**

This work was supported by the National Natural Science Foundation of China (11504041, 11574040), the Fundamental Research Funds for the Central Universities of China (DUT16LAB01, DUT17LAB19), and the Supercomputing Center of Dalian University of Technology.

**Table 1** Structural and electronic properties of the graphene/MXene heterostructures, including lattice mismatch ($\delta$), interlayer distance ($d$), vertical buckling of the graphitic sheet ($\Delta d$), interlayer binding energy per C atom in graphene ($\Delta E$), C $p_z$ band center referred to the Fermi level ($\varepsilon_{pz}$), and work function of the heterostructures ($\Phi$). The graphene sheet is doped by N atoms in the pyridinic form.

| system | $\delta$ (%) | $d$ (Å) | $\Delta d$ (Å) | $\Delta E$ (eV) | $\varepsilon_{pz}$ (eV) | $\Phi$ (Å) |
| --- | --- | --- | --- | --- | --- | --- |
| G/Ti$_2$C | 2.12 | 2.13 | 0.13 | −0.42 | −4.95 | 4.56 |
| G/V$_2$C | 1.84 | 2.08 | 0.05 | −0.32 | −5.80 | 5.31 |
| G/Nb$_2$C | 1.15 | 2.40 | 0.28 | −0.25 | −5.01 | 4.65 |
| G/Mo$_2$C | 0.23 | 2.39 | 0.27 | −0.23 | −5.12 | 4.70 |

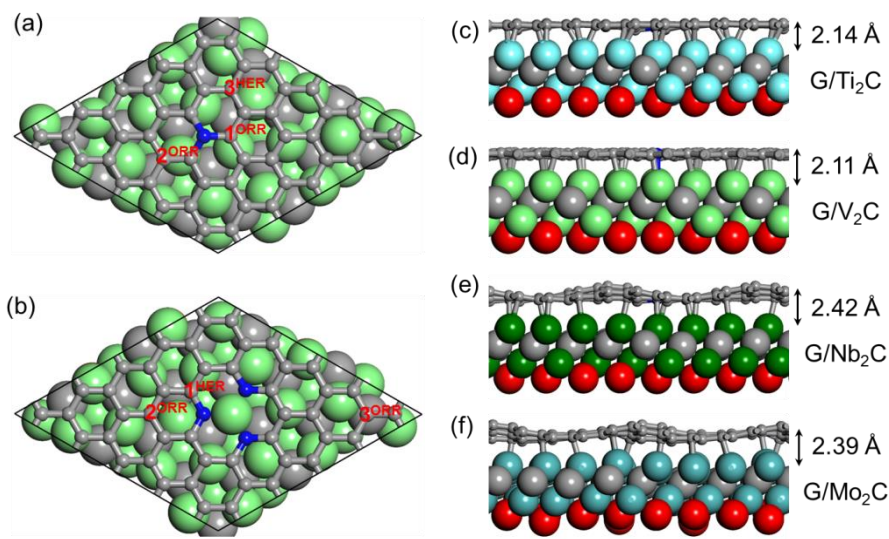

**Fig. 1** (a, b) Top views of N-doped graphene supported by V$_2$C MXene monolayer, with N dopants in the graphitic and pyridinic forms, respectively. The red numbers indicate the active sites for ORR and HER. (c, d, e, f) Side views of N-doped graphene on Ti$_2$C, V$_2$C, Nb$_2$C and Mo$_2$C MXene monolayers, respectively. The interlayer distances are shown for each system. The C, N, O, Ti, V, Nb, and Mo atoms are shown in grey, blue, red, cyan, green, dark green and turquoise colors, respectively.

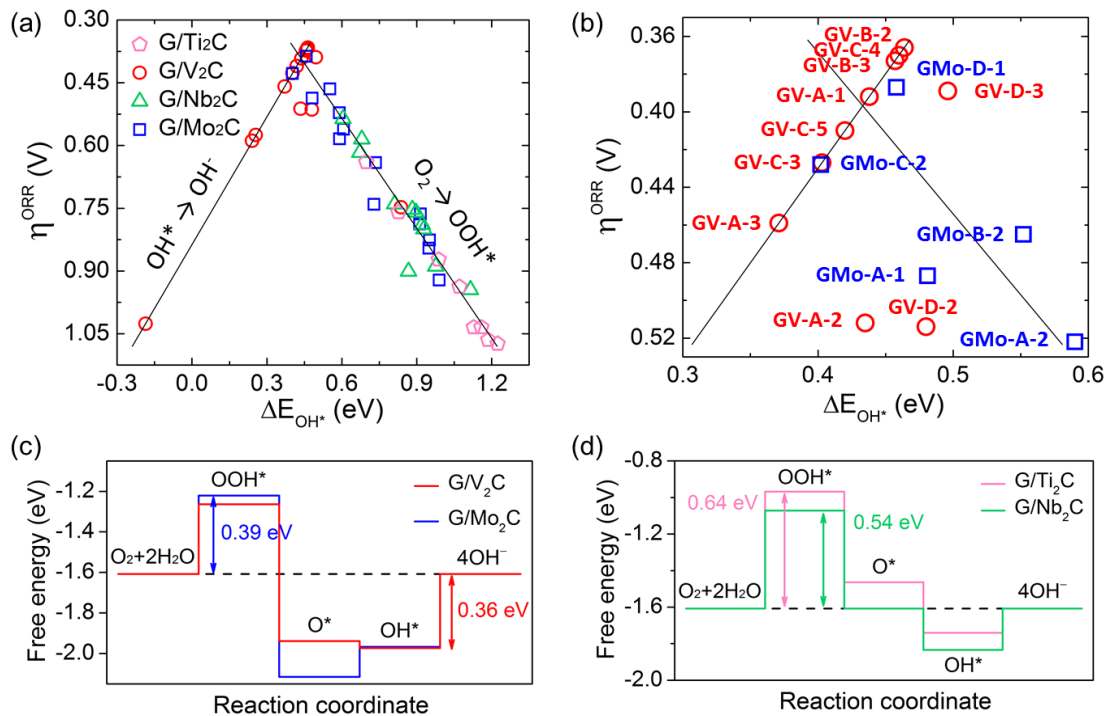

**Fig. 2** (a) Volcano plots of ORR overpotential vs. OH* binding energy and (b) the zoom-in plot close to the summit of the volcano. "GV" and "GMo" are abbreviations for G/V$_2$C and G/Mo$_2$C, respectively. (c, d) Free energy diagrams of ORR in the alkaline media (pH = 14 and $T$ = 298 K) at the equilibrium potential ($U_0$ = 0.40 V) for ideal catalyst (black dashed lines) and the most active sites of the graphene/MXene heterostructures (colored solid lines). The colored arrows and the numbers next to them indicate the rate-limit steps and the ORR overpotentials, respectively.

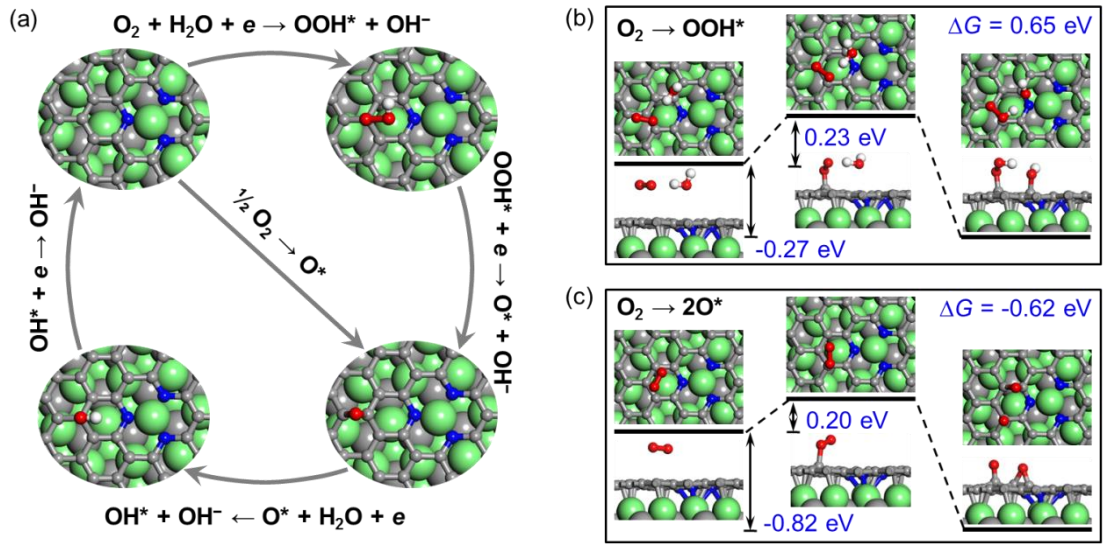

**Fig. 3** (a) Dual reaction pathways for ORR on N-doped graphene on $V_2C$ MXene monolayer. (b, c) Kinetic barriers and transition states (middle panel) for $O_2$ dissociation via two reaction pathways. The blue numbers (from left to right) indicate the total energy change during the reaction, kinetic barrier and Gibbs free energy of formation, respectively. The C, N, O and V atoms are shown in grey, blue, red and green colors, respectively.

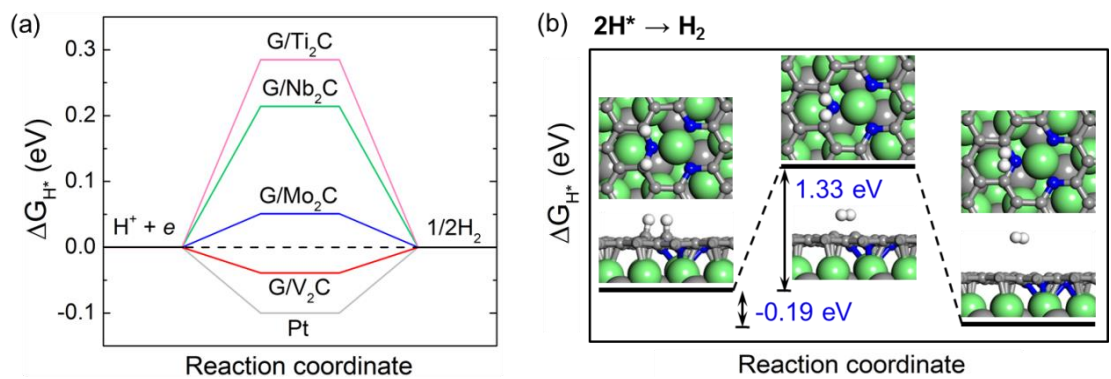

**Fig. 4** (a) Reaction free energy ($\Delta G_{H^*}$) for HER on the most active sites of various graphene/MXene heterostructures and on Pt(111) surface. (b) Change of $\Delta G_{H^*}$ during HER on N-doped graphene on $V_2C$ MXene monolayer. The insets show (from left to right) the atomic structures of initial, transition and final states, respectively. The H, C, N and V atoms are shown in white, grey, blue and green colors, respectively.

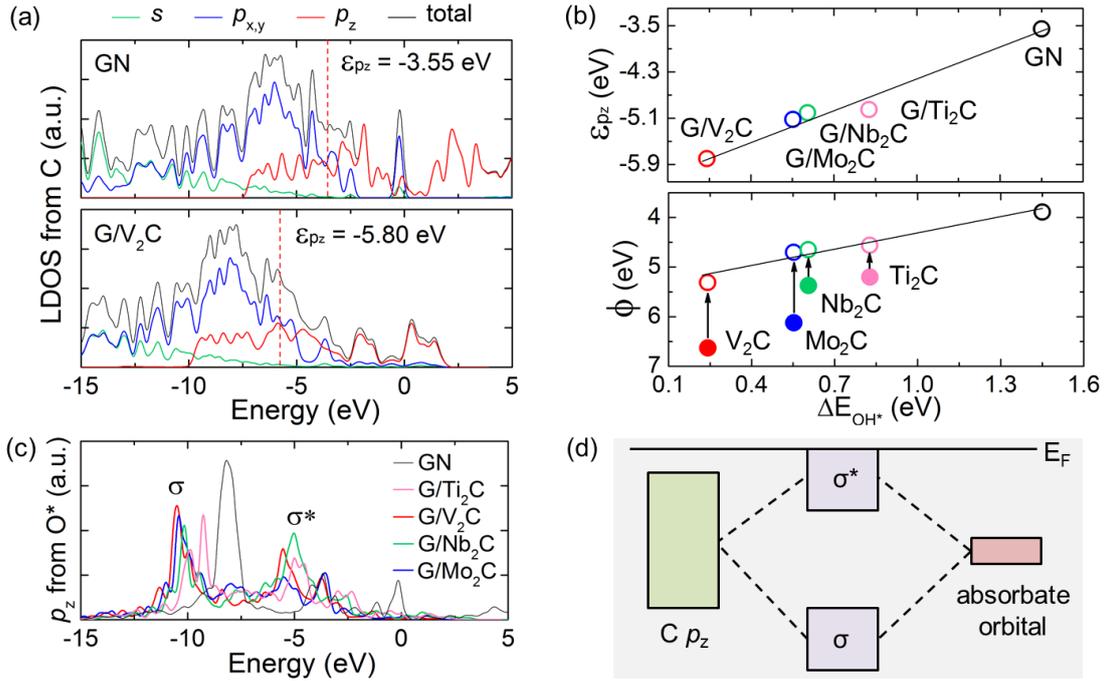

**Fig. 5** (a) Local density of states (LDOS) of the surface C atoms from freestanding N-doped graphene (top panel, abbreviated as GN) and that supported by $V_2C$ monolayer (bottom panel). The colored lines show the projected DOS from different atomic orbitals. The red dashed lines and the numbers next to them indicate the $p_z$ band center. (b) The $p_z$ band center (top panel) and work function (bottom panel) as a function of lowest binding energies of OH* species for various graphene/MXene heterostructures (colored open circles) and for GN. The work functions of standalone O-terminated MXenes are also shown for comparison (colored solid circles). (c) DOS of the $p_z$ orbital from O* species adsorbed on various graphene/MXene heterostructures and on GN. (d) Schematic diagram of orbital hybridization between C and adsorbate atoms, forming fully filled bonding ($\sigma$) and antibonding ($\sigma^*$) orbitals. The N atoms are in the pyridinic form for the data presented in (a, b, c).